\documentclass[aps,prl,twocolumn,groupedaddress,floatfix]{revtex4}

\usepackage{graphicx}
\usepackage{natbib}
\usepackage{amsmath}

\begin{document}

\title{Manufacturing a thin wire electrostatic trap (TWIST) for ultracold polar molecules}

\author{J. Kleinert}
\author{C. Haimberger}
\author{P. J. Zabawa}
\author{N. P. Bigelow}

\affiliation{Department of Physics and Astronomy, and The
Laboratory for Laser Energetics\\ The University of Rochester,
Rochester, NY 14627}

\date{\today}

\begin{abstract}
We present a detailed description on how to build a Thin WIre electroStatic Trap (TWIST) for ultracold polar molecules. It is the first design of an electrostatic trap that can be superimposed directly onto a magneto optical trap (MOT).
We can thus continuously produce ultracold polar molecules via photoassociation from a two species MOT and instantaneously trap them in the TWIST without the need for complex transfer schemes. Despite the spatial overlap of the TWIST and the MOT, the two traps can be operated and optimized completely independently due to the complementary nature of the utilized trapping mechanisms.

\end{abstract}

\maketitle

Electrostatic trapping is a robust technique for confining cold polar molecules in low field seeking states and has been demonstrated in several laboratories \cite{Meijer2,Rempe2}. 
To be able to apply this method to molecules formed in a magneto optical trap (MOT) and trapped in situ, strong constraints on optical access are imposed.
We solve this problem by using thin wires, which are submerged directly into the light fields of the MOT, rather than extended electrodes. The resulting electric field gradients - and hence the trap depth - using the TWIST are considerably weaker than in the established trap designs, however, these limitations are offset by the ultracold temperature of photoassociated molecules from a MOT making the TWIST a viable design for the environment of magneto optical traps.

\begin{figure}[h]
\includegraphics[width=8.5cm]{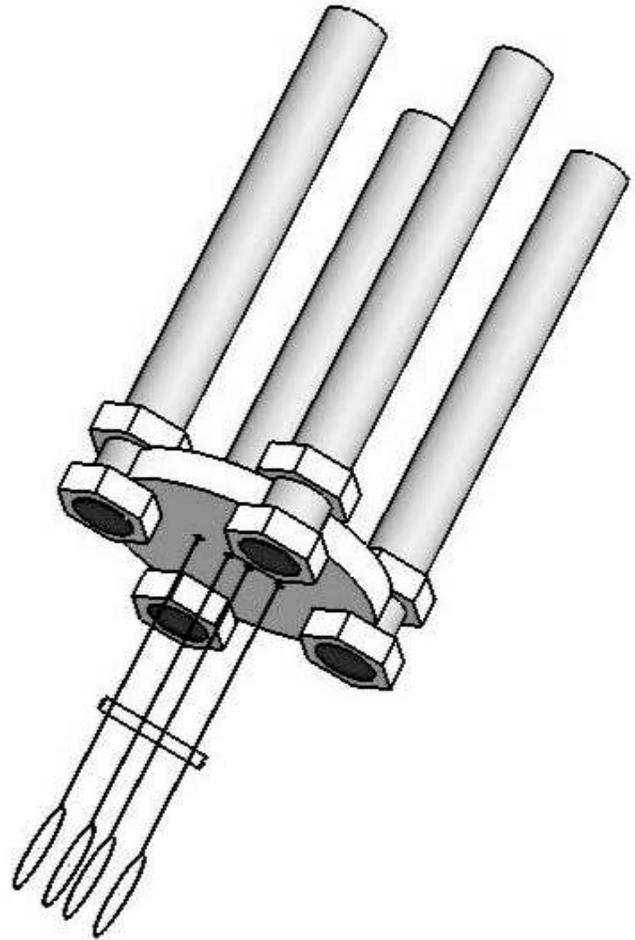}
\caption{\label{twist} Fully assembled TWIST.}
\end{figure}

The TWIST is a new design to join trapping of polar molecules with electric fields and laser cooling and trapping techniques for atoms in a unique way: first, the traps are spatially overlapped, which eliminates the need for transfer/loading schemes as the molecules are created inside the trap. Second, the respective fields of the traps do not influence each other: the atoms are not perturbed by the - comparatively - weak electric fields of the TWIST, while the magnetic fields of the MOT have no impact on molecules in their X$^1\Sigma$ state due to a lack of an appreciable magnetic moment. This enables completely independent operation, superposition and optimization of the two traps.

The geometry of our TWIST is shown in Fig. \ref{twist}. The TWIST consists of four tungsten rings of 8 mm diameter each, and separated by 3, 2 and 3 mm, respectively.
The separations of the 4 tungsten wires are maintained by an attached glass rod. The wire structure is mounted on a Macor$^{\textregistered}$ disk which is held in place by 4 copper rods. Each tungsten wire is electrically connected to one of the copper rods, which are in turn mounted on an electrical feed-through flange. This enables us to electrically address each wire individually.

The chosen thickness of the wire is a compromise: thin enough to limit perturbations to the light fields of the MOT and thick enough to create strong electric fields while maintaining the rigidity of the wires. 
We found 75 $\mu$m best suited for our application.

\begin{figure}[htb]
\includegraphics[width=7.4cm]{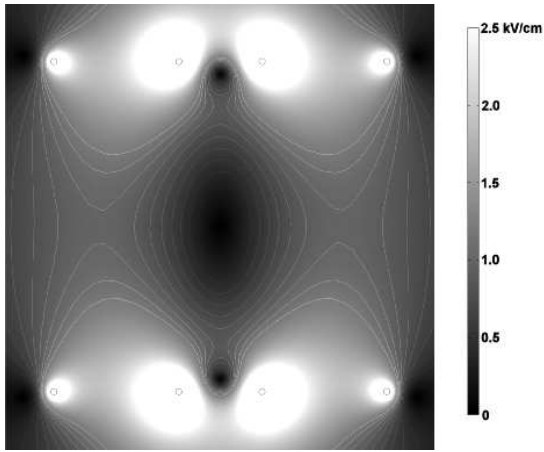}
\caption{\textit{Electric field strength contour: the outer electrode rings are grounded, the center electrodes charged to +1000 V. The ultracold polar molecules are confined in the center electric field strength minimum.}}
\label{trapfield}
\end{figure}

The electric field strength contour for our standard trapping configuration (0 V on the outer and +1000 V on the inner electrode rings) is shown in Fig. \ref{trapfield}.
While reducing the size of the outer or inner rings with respect to the other pair does result in steeper electric field gradients and hence would be advantageous for an electric field trap, an equal size of all rings has the advantage of minimizing the impact on the light fields of the magneto optical trap and also simplifies the manufacturing of the TWIST.
Reducing the size of the entire trap increases the electric field gradient as well, of course. However, we found that the impact on the number of trapped atoms in our MOT of a TWIST half the size of the one presented here (4 mm diameter electrode rings, 1.5,1 and 1.5 mm respective spacings, 50 $\mu$m wire diameter) was too detrimental to be considered a viable option. Conversely, doubling the size of the entire trap to 16 mm results in weaker electric field gradients while only marginally improving the MOT performance.

\begin{figure}[htb]
\includegraphics[width=7.4cm]{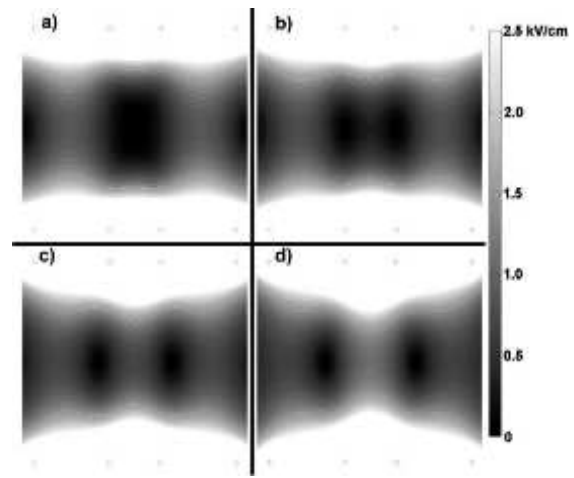}
\caption{\textit{Electric field strength contours for splitting the trapping potential. The respective voltages are adiabatically changed from (1000 V, -1000 V, 1000 V, -1000 V)$\rightarrow$( 1000 V, -200 V, 200 V, -1000 V).}}
\label{doubletrap}
\end{figure}

While only 3 electrodes would be needed to generate an electric field minimum, the 4 electrode geometry has two distinct advantages: first, it avoids an electrode at the center, which would cast a shadow right at the center of the MOT. Second, the 4 electrodes offer opportunities beyond a simple field minimum. As an example, Fig. \ref{doubletrap} demonstrates the possibility of splitting the electric field minimum in two. 

It is worth mentioning that some light, originating from the MOT lasers, scatters
off the thin wire electrodes. This can pose a problem as light with improper polarization perturbs a MOT more strongly than a mere shadow.
Therefore, the TWIST works best for Dark SPOT MOTs\cite{Pritchard2} where the dark region of the repumper is larger than the cross section of the cylinder, defined by the field generating electrodes.

\begin{figure}[htb]
\includegraphics[width=7.4cm]{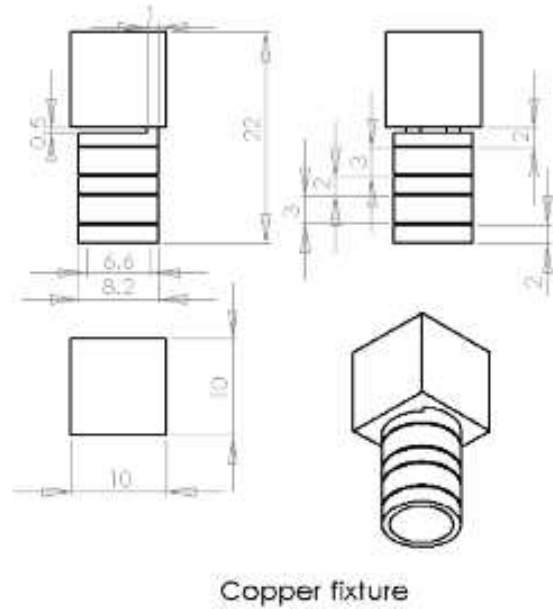}
\caption{\textit{Copper fixture}}
\label{copper}
\end{figure}

\begin{figure}[htb]
\includegraphics[width=7.4cm]{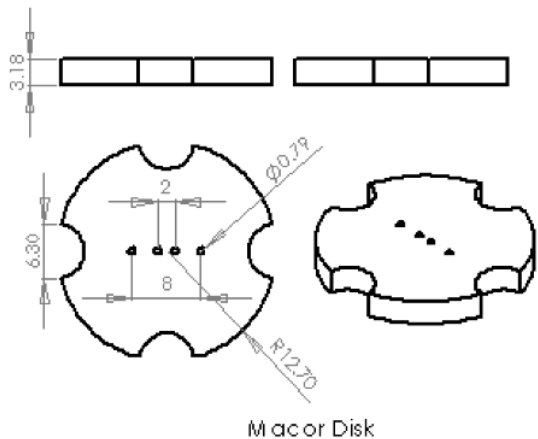}
\caption{\textit{Macor disk}}
\label{macor}
\end{figure}

\begin{figure}[htb]
\includegraphics[width=7.4cm]{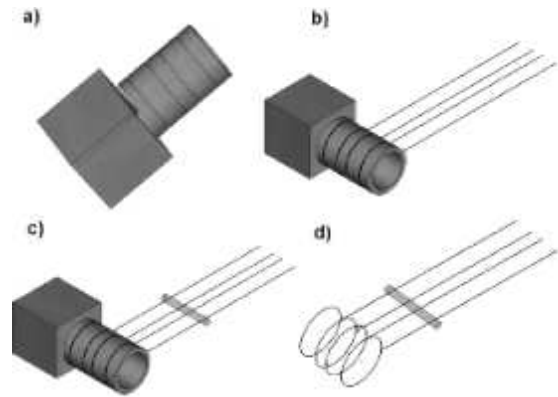}
\caption{\textit{Creating the wire structure \textbf{a)} copper fixture, \textbf{b)} copper fixture with wires, \textbf{c)} copper fixture with wires and glass rod, \textbf{d)} remaining wires and glass rod after etching.}}
\label{pieces}
\end{figure}

\begin{figure}[htb]
\includegraphics[width=7.4cm]{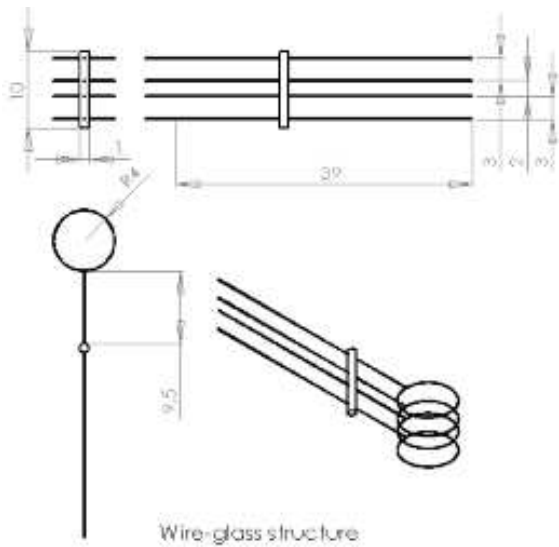}
\caption{\textit{Wire-Glass structure}}
\label{wire-glass}
\end{figure}

The TWIST wire structure is manufactured via the following steps:
\begin{itemize}
\item A copper fixture (Fig. \ref{pieces}a \& \ref{copper}), the glass rod and the Macor disk (Fig. \ref{macor}) are fabricated. 
\item The tungsten wires are wrapped around the copper fixture (Fig. \ref{pieces}b). 
\item The glass rod is attached to the tungsten wires in an inert atmosphere (Fig. \ref{pieces}c). 
\item The copper fixture is removed via etching under a fume hood, which leaves the desired bare wire structure (Fig. \ref{pieces}d \& \ref{wire-glass}). 
\item The pieces are assembled on an electric feed-through flange.
\end{itemize}
Each of these steps is described in detail in the following paragraphs.

We start by making a copper fixture (Fig. \ref{copper}). It has four indentations for the proper spacings of the rings (3, 2 and 3 mm, respectively) and an outer diameter of 8 mm corresponding to the intended size of the future wire structure. The cylindrical part of the fixture is only connected via a thin bridge to the square mounting part. This removes the need to saw the cylinder off after the glass rod has been attached to the wires; instead it is possible to just break it off. This eliminates the risk of damaging the combined wire-glass structure due to vibrations inherent in the sawing process. The cylindrical part of the copper fixture is hollow to make the etching process faster.

The use of the glass rod will ensure that the proper spacing of the wire rings is maintained. 
We choose low temperature melting glass that is commonly used in glassworking\footnote{"soft glass", Coefficient of expansion: 104}. A simple MAPP gas  or even propane gas torch is sufficient to melt the glass.
Glass rods with varying diameters are easy to produce after a few minutes of practice. We found rod diameters of 0.75-1.0 mm best suited: thicker rods are increasingly harder to attach reliably to the tungsten wire without subsequent shattering due to thermal stress,  thinner rods break too easily under mechanical stress.

Further, we need the Macor disk (1" diameter, 1/8" thickness) that we will use later on to connect the wire-glass structure to the 4 copper rods (see Fig. \ref{twist}).
Macor is a UHV compatible ceramic with excellent machining characteristics, provided carbide tools are used.
The disk is shown in Fig. \ref{macor}.: Four 1/32" holes are drilled symmetrically through the center with the distances of 3, 2 and 3 mm, respectively, corresponding to the intended respective distances of the four tungsten wires. Four partial holes are drilled on the edges to be able to attach the 4 copper rods via pairs of nuts once the whole structure is assembled. The outer 4 partial holes are positioned to have an angle with respect to the 4 center holes, such that the center holes are precisely aligned to the z-Axis of the MOT after mounting the structure on the electric feed through flange onto the chamber. If a rotating flange connection is used on the chamber side, this angle becomes arbitrary, of course.

The next part in the manufacturing process of the TWIST is the creation of the actual wire structure:
We choose tungsten wire for its stiffness, very high melting point and UHV compatibility. We form a loop with a 6" wire piece, slide it in position on the copper fixture and twist it with a hemostat. The wire is not strong enough to be twisted all the way; once the tear drop shrinks to about 1 cm, we grab the edge of the tear drop with the hemostat and twist until the tear drop is less than 0.5 mm in size. To protect the wire from any punctual stress at the hemostat's tips and edges, we keep a piece of paper between the hemostat and the wire. Once finished, the loop sits snugly on the copper piece.
We repeat this for all four wire loops (Fig. \ref{pieces}b). Wires that got bent in the twisting process will be straightened later.

Next, the glass rod is attached to the tungsten wires. To get a reliable connection between the glass and the tungsten wires, we found it not sufficient to heat the glass rod and put it on the cold tungsten wires. To be able to heat the tungsten wires sufficiently without any oxidization of the tungsten, we move the parts into an inert atmosphere glove box.
Inside, the tungsten wires can be heated by running an electric current through them until they are glowing white, just as in an incandescent light bulb. If the wires burn during heating or become very brittle, the inert atmosphere is contaminated.
 We feed the four wires that are mounted on the copper fixture through the four holes of the Macor disk. This ensures the proper respective spacing of the wires over the full length once they are straight and parallel.
To straighten any bent wires and to attach the glass rod to the wires, we heat the tungsten wires one by one by running an electric current of $\approx 1.8$ A through them:
The copper fixture is grounded. We use a metal tweezer to gently pull one of the wires and then touch the tweezer with a current probe completing the electric circuit. The indirect connection of the current probe to the wire is important, as it prevents the possible welding of the current probe to the wire via a spark. We increase the current flow until the wire is glowing white. Any small bends and kinks straighten out now as we gently pull the glowing wire with the tweezer. We repeat this for each wire.
The glass rod is attached at the minimum distance from the trap center that will still be outside the intersection of the MOT laser beams; in our case that is 0.6".
We attach the glass rod by heating up one wire after the other in the same fashion that was previously employed to straighten them. To avoid overheating and subsequent melting of the tungsten wire in this process, we implement the following procedure: we start with a low current ($\sim$ 1 A) through the tungsten wire and slowly increase it, observing the change of the emission characteristic of the hot wire from red to orange to white and stop increasing the current further when the glass - which rests on the wire - starts to melt onto it. This is usually the case at $\sim$ 1.8-2.0 A. We then operate at this current for all four wires and let the glass melt around each wire, such that each wire is fully enclosed by glass afterwards.
Before we move the parts out of the protected atmosphere glove box, we heat up each wire briefly ($\sim$ 3 s) once again with the same current used to melt the glass onto it and slowly ($\sim$ 10 s) ramp down the current and thus the temperature to anneal the tungsten.

To remove the copper fixture and obtain the bare wire-glass structure (Fig. \ref{wire-glass}), we break off the square mounting part and submerge the remaining copper cylinder with the wires in nitric acid. Due to the emission of NO$_{X}$ gases during the dissolving process of the copper it is necessary to do this under a fume hood. The tungsten wire is not affected by the nitric acid if it is only kept in the bath until the copper is dissolved.
Once the copper is completely dissolved, the wire-glass structure has its final form. If it distorted from the intended shape after removing the copper fixture, the wires were not annealed properly in the process of connecting them to the glass rod and the previous steps have to be repeated from the start.
If the wire-glass structure remains intact, it can be cleaned in an ultrasonic sound bath to prepare it for installation in an UHV chamber.
A note of caution: the wires are so thin that the surface tension of the various solvents can put considerable strain on the structure and even cause it to break, especially if the tungsten wires partially oxidized and therefore became brittle due to a slightly contaminated inert atmosphere.  Hence we remove it wire tips first from any liquid as a precautionary measure.

The final assembly takes place the following way:
First we put the top part together, i.e. the copper rods attached to the electric feed-through flange and the Macor disk to the copper rods (see Fig. \ref{twist}).
Then we attach and align the cleaned wire-glass structure to the Macor disk with hooks. These hooks consist of short pieces of tungsten wire and connect one of the four copper rods with one of the four wires each, bending it slightly and thus fixing it to the Macor disk. Once all four hooks are in place, connected and fixed on one end by tightening the respective nuts (see Fig. \ref{twist}), the fine alignment is done by pulling and pushing single wires until the assembly is straight and has the proper total length.

The assembly is then placed into the vacuum chamber.
We ensure proper electrical contact of each wire by applying a high voltage to one feedthrough at a time while grounding the others: at about 2kV each wire loop starts to bend with respect to the other wires by about one wire diameter.

Baking the wire-glass structure is challenging due to its very weak thermal connection to the vacuum chamber body.
A stab-in heater, a halogen lamp mounted on an electric feed-through flange, overcomes this. 

\begin{figure}[htb]
\includegraphics[width=7.4cm]{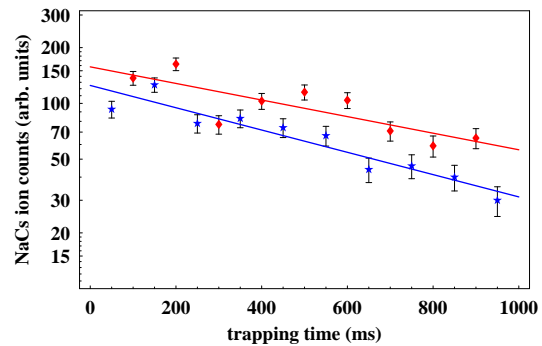}
\caption{\textit{Two measurements of the lifetime of trapped NaCs molecules in the TWIST: 850$^{+190}_{-105}$ms}}
\label{lifetime}
\end{figure}
Fig. \ref{lifetime} demonstrates the successful implementation of the TWIST into a two-species MOT apparatus. It shows the number of trapped NaCs molecules in the TWIST as a function of time at a pressure of $6.5\cdot10^{-10}$ Torr. 
Two spatially overlapped magneto-optical traps inside the TWIST cool and confine sodium and cesium atoms to temperatures of $\sim$200 $\mu$K. 
The NaCs molecules are created via a single step photoassociation from the ultracold atoms and subsequent spontaneous decay into deeply bound X$^1\Sigma$ levels. 
The chosen free bound photoassociation transition in this experiment is a $\Omega'$=0 J'=3 state, $\sim$23 GHz below the cesium D2 line. 
As the atom clouds are inside the TWIST, the molecules are created in a trapped state, which enables the continuous accumulation of NaCs in the trap. 
After an accumulation time of one second, the light of the magneto optical traps is switched off. 
A pure sample of trapped NaCs molecules remains in the detection region after 20 ms, as the atoms drift out due to thermal expansion and gravity, while the molecules remain trapped in the field of the TWIST. 
After a varying delay time (50-950 ms) - the trapping time - the high voltage of the TWIST electrodes is switched off and a 500 $\mu$J, 10ns pulse of light at 591.3 nm ionizes the NaCs molecules, which are subsequently detected in a channel electron multiplier (CEM).
We determined the lifetime to be 850$^{+190}_{-105}$ ms, a significant improvement compared to a previous measurement of 225$\pm30$ms at $4\cdot10^{-9}$ Torr \cite{Bigelow4}.

\begin{acknowledgments}
We would like to thank C. Forrest for his help in planning and building a first prototype of the TWIST and Prof. Novotny for providing a Coherent Verdi.
This work was supported by the National Science Foundation and The Army Research Office.  Chris Haimberger and Jan Kleinert are grateful to the Laboratory for Laser Energetics for DOE Horton Fellowships.
\end{acknowledgments}

\end{document}